\documentclass[prl,twocolumn,showpacs,preprintnumbers,amsmath,amssymb]
{revtex4}

\usepackage{graphicx}

\begin{document}

\title{Recursion method for the quasiparticle structure of a single vortex
    with induced magnetic order}
\author{Linda Udby$^1$, Brian M. Andersen$^2$, and Per Hedeg\aa rd$^3$}
\affiliation{$^1$Materials Science Dept., Ris{\o}   National
Lab., Frederiksborgvej 399, DK-4000 Roskilde, Denmark\\
$^2$Department of Physics, University of Florida, Gainesville,
Florida 32611-8440, USA\\ $^3$\O rsted Laboratory, Niels Bohr
Institute, Universitetsparken 5, DK-2100 Copenhagen \O, Denmark}

\date{\today}

\begin{abstract}
We use a real-space recursion method to calculate the local density
of states (LDOS) within a model that contains both $d$-wave
superconducting and antiferromagnetic order. We focus on the LDOS in
the superconducting phase near single vortices with either normal or
antiferromagnetic cores. Furthermore, we study the low-energy
quasiparticle structure when magnetic vortices operate as pinning
centers for surrounding unidirectional spin density waves (stripes).
We calculate the Fourier transformed LDOS and show how the energy
dependence of relevant Fourier components can be used to determine
the nature of the magnetic field-induced order, and predict
field-induced LDOS features that can be tested by future scanning
tunneling microscopy (STM) experiments.
\end{abstract}


\maketitle

\section{Introduction}
It is becoming evident that competing phases cause many of the
anomalous properties of doped Mott insulators. An example is given
by the vortex state of underdoped high-$T_c$ superconductors where
antiferromagnetism (AF) 'pops up' near the
vortices\cite{zhang,arovas}. Initial experimental evidence for this
claim came from STM experiments on YBa$_2$Cu$_3$O$_{y}$ (YBCO) and
Bi$_2$Sr$_2$CaCu$_2$O$_{8+x}$ (BSCCO) observing weak low-energy
quasiparticle peaks around 5-7 meV\cite{maggio,pan}. This strongly
contradicts the expected LDOS in the vortex center of a pure BCS
$d$-wave superconductor (dSC) which is dominated by the so-called
zero-energy state (ZES), a single broad resonance centered at the
Fermi level\cite{wang}. Further evidence for AF cores has come from
both nuclear magnetic resonance measurements\cite{mitrovic} and muon
spin rotation experiments\cite{muonmiller}. The field-induced
magnetization is not necessarily restricted to the core regions as
determined by the coherence length $\xi$. For instance, elastic
neutron scattering on underdoped La$_{2-x}$Sr$_{x}$CuO$_{2}$ (LSCO)
showed that the intensity of the incommensurate peaks in the
superconducting phase is considerably increased when a magnetic
field is applied perpendicular to the CuO$_{2}$ planes\cite{bella}
or when Zn is doped into the samples\cite{Zn}. Similar results have
been found in the oxygen doped sample
La$_{2}$CuO$_{4+y}$\cite{khaykovich}. The momentum position and
field-enhanced sharpening of this elastic signal corresponds to a
spin density wave period of roughly eight lattice constants $8a$
extending far outside the vortex cores, suggesting that the magnetic
cores operate as pinning centers for surrounding spin density
waves\cite{demler,andersenpinned1}. This unusual behavior agrees
with in-field STM measurements on optimally doped BSCCO which found
local field-induced checkerboard LDOS patterns with a period close
to $4a$\cite{hoffman}. Similar structure has been reported in zero
field STM experiments\cite{howald}. Pronounced checkerboard ordering
has also been detected in
Na$_x$Ca$_{2-x}$CuO$_2$Cl$_2$\cite{hanaguri}. More recently, Levy
{\sl et al.}\cite{levy} confirmed the results of Ref.
\onlinecite{hoffman} and found that the checkerboard modulation does
not disperse with energy, and mapped out the energy dependence of
the amplitude of the Fourier component corresponding to the ordering
vector of the modulation.

Theoretically, several groups have proposed that the origin of the
unexpected behavior inside the cores is related to locally nucleated
AF\cite{afother,knapp}, but other scenarios have also been
proposed\cite{otherscenarios,franztesanovic}. From a computational
point of view, in order to model the existence of nano-scale
inhomogeneity, it is necessary to use methods that easily allows one
to obtain the LDOS as a function of energy and large real-space
regions. Traditionally this is done by numerical diagonalization of
the Bogoliubov-de Gennes (BdG) equations, which, at present, is
typically restricted to quite small lattices ($ \lesssim 40\times40$
sites). In this paper we use a recursion method generalized to the
$d$-wave superconducting state to calculate the LDOS near an
increasingly complex single vortex. This method is easily applied to
large systems allowing for e.g. high-resolution Fourier LDOS images.
First we study the pure dSC vortex for realistic band structure
parameters relevant for overdoped cuprates. Second, we discuss the
case of an AF vortex core in the optimally doped regime and focus on
the spatial dependence of the expected LDOS. Finally, we calculate
the LDOS when the vortex pins surrounding incommensurate stripe
order as may be relevant for LSCO and underdoped BSCCO, and discuss
the energy dependence of the resulting Fourier transform. As opposed
to most earlier theoretical work on the AF vortex
problem\cite{afother}, we focus on the final LDOS structure and the
Fourier transformed LDOS maps which can be used as an STM tool to
determine the nature of the field-induced order and the origin of
the ZES splitting. Lastly, we compute the LDOS resulting from the
recently proposed pair-density wave ordered state.

\section{Model and Method}

In the following we study the mean-field Hamiltonian defined on a 2D lattice
\begin{eqnarray}\label{hamil}
\hat{H} =- \sum_{\left< ij
  \right>\sigma}t_{ij}\hat{c}^\dagger_{i\sigma} \hat{c}_{j\sigma} &-& \mu \sum_{i\sigma}
\hat{c}^\dagger_{i\sigma} \hat{c}_{i\sigma} \\ \nonumber
+ \sum_{\left< ij \right>} \left( \Delta_{ij}
\hat{c}^\dagger_{i\uparrow} \hat{c}^\dagger_{j\downarrow} +\mbox{H.c.}\right)
&+& \sum_{i} m_i
\left(\hat{c}^\dagger_{i\uparrow} \hat{c}_{i\uparrow} - \hat{c}^\dagger_{i\downarrow}
  \hat{c}_{i\downarrow} \right),
\end{eqnarray}
where $\hat{c}^\dagger_{i\sigma}$ creates an electron with spin
$\sigma$ at site $i$, $t_{ij}$ is the hopping integral to nearest
($t$) or next-nearest ($t'$) neighbors and $\mu$ is the chemical
potential. The AF and dSC order parameters are given by $m_i$ and
$\Delta_{ij}$, respectively. The Hamiltonian (\ref{hamil}) is the
effective mean-field model obtained after performing two
Hubbard-Stratonovich transformations of the extended Hubbard model
with the on-site repulsion causing the AF, and the attractive
nearest neighbor interaction resulting in the dSC. It has been used
extensively in the past few years to gain insight into the
electronic structure of phases of coexisting AF and dSC
order\cite{andersenpinned1,afother,andersenpinned2,granath,yuripaper}.

Below, we solve the Hamiltonian (\ref{hamil}) using appropriate
Ans\"{a}tze for both $\Delta_{ij}$ and $m_i$. The lack of
self-consistency can sometimes be useful in clarifying, for
instance, the nature/origin of vortex core states\cite{berthod}. We
restrict the discussion to the case when the applied magnetic field
is much smaller than $H_{c2}$ and consequently ignore the vector
potential ${\mathbf{A}}$.

In order to obtain the LDOS near single vortices we use a recursion
method\cite{haydock,miller} generalized to the superconducting
state. The starting point is to generate a new orthonormal basis of
states from the recursion relation
\begin{equation}
\hat{H}\left|n\right>=a_n\left|n\right>+b_{n+1}\left|n+1\right>+b_n\left|n-1\right>.
\end{equation}
For each recursion the Greens function of the $n$'th level is
generated recursively from the Lanczos coefficients $a_n$ and $b_n$
\begin{equation}
G_n(\omega)=\frac{1}{\omega-a_n-b^2_{n+1}G_{n+1}(\omega)}.
\end{equation}
Hence, the local Greens function can be found if $G_N(\omega)=0$ for
some number $N$, or if an appropriate analytical solution of $G_N$
for an infinite chain can be attached.

The retarded Greens function is
\begin{equation}
G^R_{i\sigma}(\omega) \!=\!\!\sum_\alpha \left(
\frac{\big|\left<\alpha\right|c_{i\sigma}^\dagger
\left|0\right>\big|^2}{\omega-E_{\alpha
\sigma}+i\eta}\!+\!\frac{\big|\left<\alpha\right|c_{i\sigma}
\left|0\right>\big|^2}{\omega+E_{\alpha\sigma}+i\eta}\right),
\end{equation}
where $\eta$ is used as an artificial smearing factor with $\eta
=0.02t$. In general, it is not necessary to perform four recursions
($I$ from $c_{i\uparrow}^\dagger|0\rangle$; $II$ from
$c_{i\downarrow}|0\rangle$; $III$ from
$c_{i\downarrow}^\dagger|0\rangle$; $IV$ from
$c_{i\uparrow}|0\rangle$) to calculate the spin-summed LDOS since
$G^{IV}_{n=0}(-\omega)= G^{I}_{n=0}(\omega)$ and
$G^{III}_{n=0}(\omega)= G^{II}_{n=0}(-\omega)$. Thus, it is
sufficient to perform only two recursions to obtain the total LDOS,
$\rho(i,\omega)$, in the form of a continued fraction
\begin{eqnarray}\nonumber
\rho(i,\omega)
&=&\mbox{Im}\frac{-\frac{1}{\pi}}{\omega-a_0^{I}+i\eta-\frac{(b_1^{I})^2}
{\omega-a_1^{I}+i\eta-\frac{(b_2^{I})^2}{\omega-a_2^{I}+i\eta-\ldots}}}\\
& &+(\omega \rightarrow -\omega, a^I \rightarrow a^{II}, (b^I)^2
\rightarrow (b^{II})^2),
\end{eqnarray}
which can be compared to the differential tunneling conductance as
measured by e.g. an STM tip. Of course, when there is spin
degeneracy (here: $m_i=0$) only one recursion is needed to produce
the total LDOS.

In the cases studied, we find that the Lanczos coefficients converge
nicely when increasing the number of recursions, i.e. the system
size. Below we simply perform the truncation $G_N=0$ where $N$ is
some number of order $10^3$, and have checked that this choice does
not affect the reported results.

\section{Results}

In this section we use the recursion method to study
$\rho(i,\omega)$ around a single vortex in the dSC state both with
and without antiferromagnetism in the core region. This is supposed
to model the vortex LDOS in the overdoped and optimally doped
regime, respectively. The core center is positioned at the origin
$(0,0)$ and lengths are measured in units of the lattice constant
$a$.

\subsection{A single vortex without induced stripe order}

As is well-known, in an $s$-wave BCS superconductor the vortex
generates states localized transverse to the flux line. These have
been studied in great detail both
theoretically\cite{caroli,gygi,shore} and experimentally\cite{hess}.
The core states result from the opposite sign of the supercurrent
term in the particle and hole part of the BdG
equations\cite{caroli,berthod,schopohl}. The reduction of the pair
potential near the vortex core causes only minor quantitative
changes to these states. We have verified that the recursion method
described above applied to $s$-wave superconductors successfully
reproduce these Caroli-de Gennes-Matricon bound states.

To model an isolated $d$-wave vortex the following pairing potential
is used
\begin{equation}
\Delta_{ij}=\Delta \tanh(|{\mathbf{r}}|/\xi) \exp(i \varphi_{ij}),
\end{equation}
where $\Delta$ is positive (negative) on x (y) links,
${\mathbf{r}}=({\mathbf{r}}_i+{\mathbf{r}}_j)/2$ and $\exp(i
\varphi_{ij})=(x+iy)/r$ with ${\mathbf{r}}=(x,y)$. In agreement with
Ref. \onlinecite{berthod}-\onlinecite{schopohl}, we find that the
suppression of the gap in the core region results in only minor
quantitative changes: in general the suppression tends to push the
states slightly further toward the Fermi level. In the pure dSC
state, the vortex is dominated by the well-known ZES\cite{wang}.
However, the ZES is centered exactly at zero energy only for
$\mu=t'=0$. As opposed to the Caroli-de Gennes-Matricon states in
the $s$-wave vortex, the ZES is made up of several states that merge
to form the broad peak as the system size is increased in agreement
with the extended nature of this peak\cite{franztesanovic}.
\begin{figure}[t]
\includegraphics[width=8.0cm]{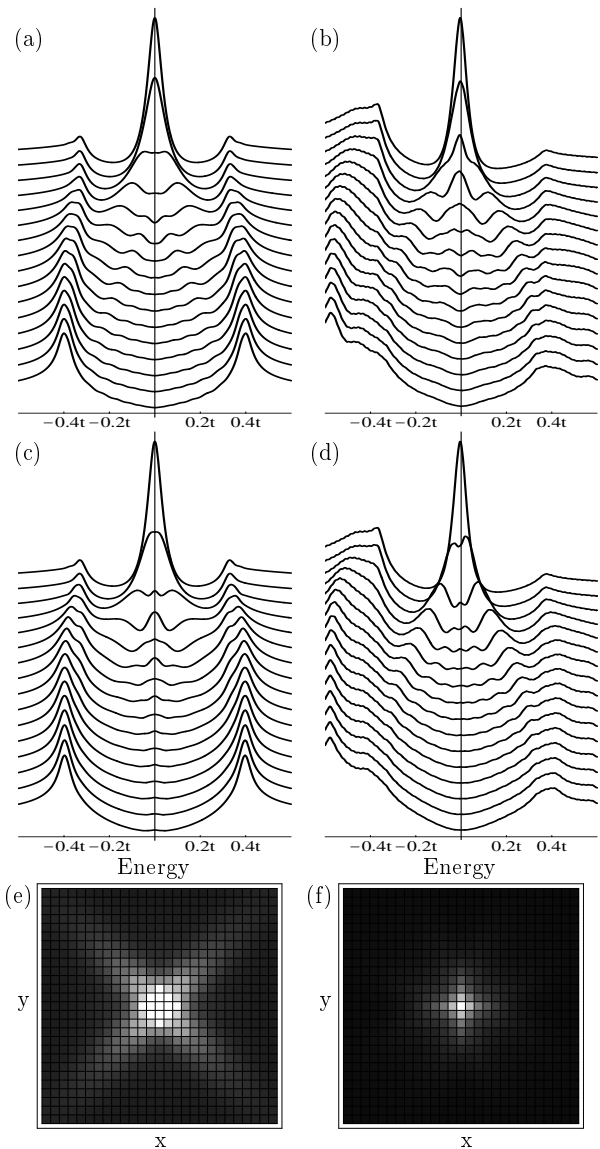}
\caption{LDOS along the anti-nodal (a-b) and nodal (c-d) direction
for a single dSC vortex. (e-f) Spatial 2D structure of the ZES. Left
column: $\mu=t'=0$, $\Delta=0.1t$. Right column: $\mu=-1.18t$,
$t'=-0.4t$, and $\Delta=0.1t$.} \label{dvortex}
\end{figure}
In Fig. \ref{dvortex} we show $\rho(i,\omega)$ of a dSC vortex along
the anti-nodal (a-b) and nodal (c-d) directions for $\Delta=0.1t$,
$\xi=5$ and $\mu=t'=0$ (a,c,e), and $\mu=-1.18t$, $t'=-0.4t$
(b,d,f). The latter parameter set provides a reasonable fit to the
Fermi surface of slightly overdoped BSCCO with a van Hove
singularity at $\omega_{vH}=-\sqrt{(4t'-\mu)^2+(4 \Delta)^2}$. It is
the $d$-wave symmetry that causes the angular dependence (compare
e.g. Fig. \ref{dvortex}(a) and \ref{dvortex}(c)) of the higher
energy core states at ${\mathbf{r}} \neq 0$. The energy and
amplitude of these core states are seen to be sensitive to the band
parameters. This is also true for the ZES state as seen from Fig.
\ref{dvortex}(e-f): at $t'=0$ the low-energy spatial form of the
LDOS has a star-shape due to the nodal dSC
phase\cite{wang,soininen,ichioka}. However, for the more realistic
BSCCO band parameters, the star is rotated with small maximum
intensity along the anti-nodal directions, which is our prediction
for the overdoped regime of BSCCO where competing AF order is
expected to be absent. For $t'=0$, it is well-known that a similar
$\pi/4$ rotation takes place at higher energies revealing the
spatial form of the higher energy core
states\cite{ichioka,zhutingbalatsky}.

\begin{figure}[b]
\includegraphics[width=8.0cm]{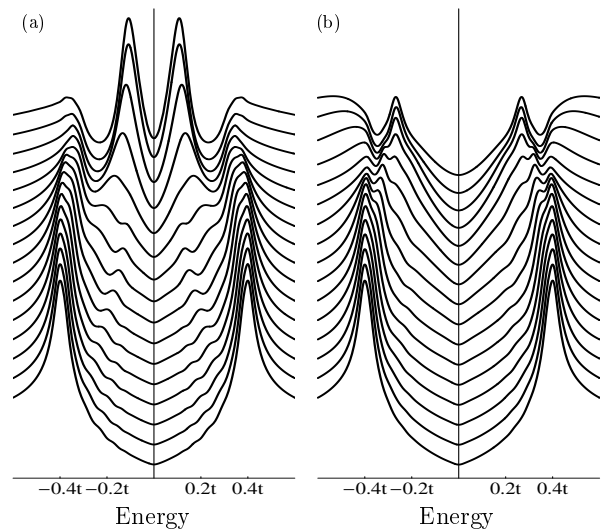}
\caption{LDOS along the anti-nodal direction for
  a dSC vortex with antiferromagnetic core. Parameters used:
  $\mu=t'=0$, $\Delta=0.1t$, $m=0.2t$ (a) and $m=0.5t$
  (b).}
\label{afvortex}
\end{figure}
We turn now to the simplest AF core situation where the suppression
of the dSC gap inside the core causes a concomitant increase of the
competing AF order\cite{arovas}. For simplicity, we model the AF
core by
\begin{equation}
m_i=m (-1)^{(x+y)} (1-\tanh(|{\mathbf{r}}_i|/\xi)),
\end{equation}
where ${\mathbf{r}}_i=(x,y)$. Due to an associated local increase of
the electron density, such vortices will in general be
charged\cite{chenting}, and have been shown to remain stable when
including the long-range Coulomb repulsion\cite{knapp}. Below, we
therefore use $\mu=t'=0.0$ in order to model the close to
half-filled vortex core regions as found in the self-consistent
studies\cite{afother,knapp}. In Fig. \ref{afvortex} we show the
final LDOS as a function of energy and distance to the AF vortex
core. The induced magnetization leads to a splitting of the ZES as
found previously\cite{afother,knapp}. With increased magnetic order
$m$, the resonant core states are pushed to higher energies and lose
spectral weight. The vortex region is fully void of apparent core
states for $m \gtrsim t$. In this limit the low-energy LDOS has an
apparent similarity with that of the pseudo-gap.

\begin{figure}[t]
\includegraphics[width=8.0cm,height=6.0cm]{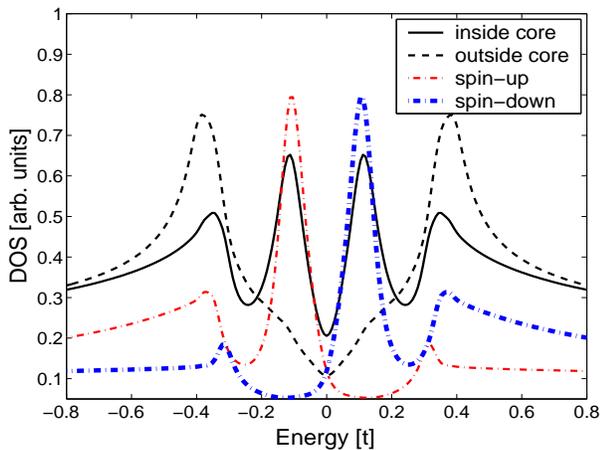}
\caption{(Color online) DOS averaged over a coherence length $\xi$
within the core (solid black line), and just outside the core
(dashed black line). The red dash-dotted line (blue thick
dash-dotted line) show the spin-up (spin-down) resolved LDOS at the
core center similar to the top scan in Fig. \ref{afvortex}(a).}
\label{AFcoresummary}
\end{figure}

Spatial averaging may mask the observability of the dispersive core
states in Fig. \ref{afvortex}. For example, in Fig.
\ref{AFcoresummary} we show $\rho(i,\omega)$ (same parameters as in
Fig. \ref{afvortex}(a)) at $(0,0)$ and $(0,7.5)$ averaged over a
coherence length $\xi$. As seen, the resulting LDOS appears to be
that of approximately {\it non-dispersing} resonant states which
rapidly lose weight when moving away from the core region. This is
similar to the measured differential tunneling conductance near the
vortex cores of YBCO and BSCCO\cite{maggio,pan}. However, an
unambiguous experimental determination of the type of order that
induces the splitting of the ZES is important, and is related to the
general discussion of time-reversal symmetry breaking for
zero-energy Andreev states in $d$-wave superconductors\cite{trsb}.
Of course, vortices that support AF cores would lead to a
field-induced {\it commensurate} $(\pi,\pi)$ signal in neutron
scattering. However, purely from a tunneling point of view, it is
possible to distinguish AF and e.g. $d_{x^2+y^2}+id_{xy}$ induced
order by using spin polarized STM\cite{wiesendanger}. This is also
shown in Fig. \ref{AFcoresummary} where the positive (negative) bias
peak is seen to be related to the spin-down (spin-up) LDOS,
respectively. Importantly, this bias asymmetry will alternate with
site, allowing for an unambiguous experimental test of AF order by a
magnetic STM tip scanned through the vortex core region.

\subsection{A single vortex with induced stripe order}

In this section we calculate $\rho(i,\omega)$ around a dSC vortex
which operates as a pinning center for unidirectional spin- and
charge density modulations (stripes), expected to be relevant for
STM experiments in the underdoped regime. Such inhomogeneous stripe
solutions indeed exist in a regime of intermediate AF coupling
within self-consistent mean-field models that include the
competition between AF and dSC order\cite{afother}. We also briefly
discuss the expected LDOS resulting from the recently proposed
pair-density-wave (PDW) induced order consisting of a density wave
of Cooper pairs without global phase coherence\cite{pairs,jhu}.

Whereas the previous section dealt with the details of
$\rho(i,\omega)$ inside the core, the field-induced periodic order
can most conveniently be studied in Fourier space
\begin{equation}
\rho_{{\mathbf{q}}}(\omega)=\frac{1}{N} \sum_i \rho(i,\omega)
e^{-i {\mathbf{q}}\cdot {\mathbf{r}}_i}.
\end{equation}
\begin{figure}[b]
\includegraphics[width=8.0cm]{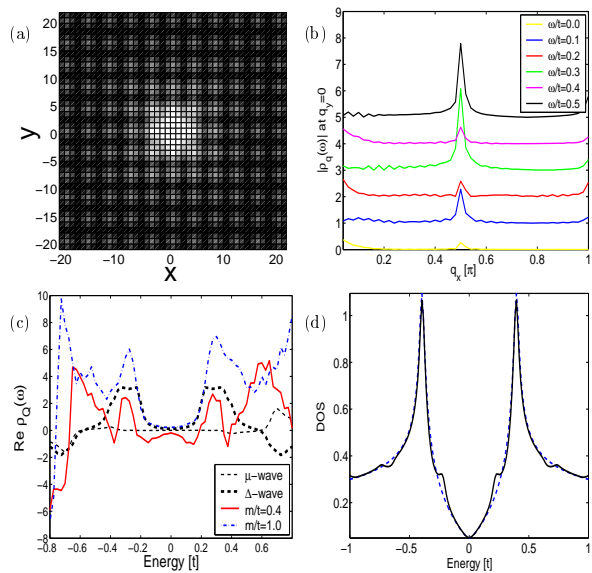}
\caption{(Color online) (a) LDOS summed in the window $\omega \in
[-0.04t,0.04t]$ with $m/t=1.0$. White (black) corresponds to high
(low) LDOS. (b) $|\rho_{\mathbf{q}}(\omega)|$ at $q_y=0.0$ vs $q_x$
for the energies: $\omega/t=0.0, 0.1, 0.2, 0.3, 0.4, 0.5$. The
curves are offset for clarity. (c) $\rho_{{\mathbf{Q}}}'(\omega)$ vs
$\omega$ for a $\mu$-wave ($\Delta$-wave, PDW) with $t'=\mu=0.0$,
$A=0.05t$ (dashed, black lines), and the full vortex induced stripe
situation with $t'=-0.4t$, $\mu=-1.18$, $m/t=0.4$ (red, solid line),
and $m/t=1.0$ (blue, dash-dotted line). Note that for all the curves
$\rho_{{\mathbf{Q}}}'(0)=0$ as expected for a $d$-wave
superconductor at $T=0$. (d) LDOS for the clean dSC (PDW) shown by
the dashed (solid) line.}\label{fig4}
\end{figure}
For site-centered stripes with spin (charge) period of 8 (4) lattice
constants, we show a typical checkerboard result for
$\rho(i,\omega)$ in Fig. \ref{fig4}(a). The checkerboard pattern
arises from including both vertical and horizontal stripes, which is
a simple way to include the assumed slow fluctuation of the stripe
domains. In addition, as shown recently, quenched disorder can
severely smear any clear distinction between stripe and checkerboard
symmetry breaking\cite{robertson}. Fig \ref{fig4}(b) shows
$|\rho_{{\mathbf{q}}}|$ as a function of $q_x$ for $q_y=0.0$ for
various energies $\omega$. The non-dispersive peak at the charge
ordering vector ${\mathbf{Q}}=(2\pi/4,0)$ resulting from the stripes
is seen to completely dominate other quasiparticle interference
effects. It is evident from Fig. \ref{fig4}(b) that
$|\rho_{{\mathbf{Q}}}(\omega)|$ displays a non-monotonic dependence
on energy. In fact, as pointed out in Ref. \onlinecite{podolsky} for
the case of {\sl weak} translational symmetry breaking, useful
information about the induced order and the underlying quasiparticle
structure is contained in $\rho_{{\mathbf{Q}}}'(\omega)$, the real
part of $\rho_{{\mathbf{q}}}(\omega)$ at the ordering vector
${\mathbf{Q}}$. In general, for weak induced order
$\rho_{{\mathbf{Q}}}'(\omega)$ will exhibit peaks near
$\omega=\omega_{vH}$ due to the logarithmic divergence coming from
the van Hove points at $(0,\pm\pi)$ and $(\pm\pi,0)$, and near
energies determined from degeneracy points
$E_{\mathbf{k}}=E_{\mathbf{k}+\mathbf{Q}}$, where $E_{\mathbf{k}}$
is the quasiparticle spectrum for the homogeneous
dSC\cite{podolsky}. For the simple nested Fermi surface $(t'=\mu=0)$
and in the case a weak unidirectional $\mu$-wave ($\mu=A\sin(2\pi/4
x)$) or a $\Delta$-wave ($\Delta=\Delta_0+A\sin(2\pi/4 x)$), this is
illustrated in Fig. \ref{fig4}(c) by the black dashed lines. For the
$\Delta$-wave ($\mu$-wave) $\rho_{{\mathbf{Q}}}'(\omega)$ is
symmetric (anti-symmetric) with characteristic peaks inside
(outside) the bulk gap as well as weight at $\omega=0.4t$ which is
the van Hove energy for this band structure\cite{pairs,podolsky}. In
Fig. \ref{fig4}(c) we also show the full numerical result for
$\rho_{{\mathbf{Q}}}'(\omega)$ in the vortex state for different
strengths of the magnetic order $m$. As seen, the stripe induced
features in $\rho_{{\mathbf{Q}}}'(\omega)$ are roughly symmetric
around $\omega=0.0$. We find that sign changes in
$\rho_{{\mathbf{Q}}}'(\omega)$ at low energy $\omega \lesssim
\Delta$ are only present for weak induced order $m/t \lesssim 0.45$.
We have checked that $\rho_{{\mathbf{Q}}}'(\omega)$ is determined
almost entirely by the stripe order: omission of the vortex flow
causes only minor quantitative changes at $\omega \lesssim \Delta$.
We expect the qualitative results presented in Fig. \ref{fig4}(c) to
apply primarily to LSCO and LBCO. In BSCCO, on the other hand, it is
becoming clear that a strong component of the LDOS inhomogeneity is
given by gap disorder\cite{podolsky,pairs,nunnerfang}.

We now turn briefly to the discussion of the LDOS near vortices with
induced PDW order which, for simplicity, is modeled with a
$\Delta$-wave. It is clear that PDW modulations cannot be the only
induced order since that would not lead to a splitting of the ZES in
the core center, and would not explain the enhanced spin response in
the neutron experiments in the mixed state. Nevertheless, the
question remains whether for certain regions of the phase diagram it
coexists with or dominates the induced spin and/or charge order
surrounding the cores, resulting in distinct features of the
measured LDOS. As shown in Fig. \ref{fig4}(c), in the case of
particle-hole symmetric bands $\rho_{{\mathbf{Q}}}(\omega)$ is a
good probe of the induced order since $\rho_{{\mathbf{Q}}}'(\omega)$
is symmetric or antisymmetric with respect to the bias voltage for
periodic modulations in the $\tau_1$ or $\tau_3$ channel of Nambu
space, respectively. However, realistic band parameters and possible
coexistence of other symmetry breakings will strongly modify
$\rho_{{\mathbf{Q}}}'(\omega)$ making detailed fitting to various
assumed order parameters necessary\cite{podolsky}. Here, we propose
the alternative possibility to search for PDW order using STM by
identifying the Andreev resonant states existing in any gap
modulated landscape\cite{nunnerfang}. In Fig \ref{fig4}(d) we show
the LDOS far away from the core region in the case where $\Delta$ is
modulated by an additional sinusoidal wave of period four and an
amplitude of $30\%$ of the average gap. As seen, the Andreev states
result in a distinct sub-gap shoulder in $\rho(i,\omega)$ {\it
inside} the bulk gap in regions well {\it outside} the vortex core.
Such sub-gap structure will be approximately non-dispersing and
hence distinct from dispersing core states extending outside the
core region.

\section{Conclusions}

We have presented theoretical results for the quasiparticle
structure near an increasingly complex vortex of a $d$-wave
superconductor. We have discussed distinct LDOS features expected
when magnetic or pair density wave order is induced by an applied
magnetic field, and have suggested new tunneling experiments to test
for field-induced antiferromagnetic order near the vortex cores of
high-$T_c$ materials.

\end{document}